\documentclass[runningheads]{llncs}
\usepackage[T1]{fontenc}
\usepackage{amsmath,amssymb}
\usepackage{booktabs}
\usepackage{multirow}
\usepackage{hyperref}

\usepackage{graphicx,verbatim}
\begin{document}
\title{Extending 2D foundational DINOv3 representations to 3D segmentation of neonatal brain MR images}
%

\author{Annayah Usman, Behraj Khan, Tahir Qasim Syed}
\authorrunning{Anonymized Author et al.}
\institute{Institute of Business Administration Karachi \\
    \email{tahirqsyed@gmail.com}}
  
\maketitle
\begin{abstract}
Precise volumetric delineation of hippocampal structures is essential for quantifying neurodevelopmental trajectories in pre-term and term infants, where subtle morphological variations may carry prognostic significance. While foundation encoders trained on large-scale visual data offer discriminative representations, their 2D formulation is a limitation with respect to the $3$D organization of brain anatomy. We propose a volumetric segmentation strategy that reconciles this tension through a structured window-based disassembly-reassembly mechanism: the global MRI volume is decomposed into non-overlapping 3D windows or sub-cubes, each processed via a separate decoding arm built upon frozen high-fidelity features, and subsequently reassembled prior to a ground-truth correspendence using a dense-prediction head. This architecture preserves constant a decoder memory footprint while forcing predictions to lie   within an anatomically consistent geometry.  Evaluated on the ALBERT dataset for hippocampal segmentation, the proposed approach achieves a Dice score of 0.65 for a single 3D window. The method demonstrates that volumetric anatomical structure could be recovered from frozen 2D foundation representations through structured compositional decoding, and offers a principled and generalizable extension for foundation models for 3D medical applications.


\end{abstract}

\section{Introduction}

The hippocampus undergoes rapid structural development during early infancy, where subtle morphological variations between term and pre-term infants can carry important prognostic information \cite{uematsu2012developmental}. Accurate volumetric segmentation of this structure is therefore critical for reliable neurodevelopmental assessment. Magnetic resonance imaging (MRI) provides high soft-tissue contrast for the developing brain and has become the preferred noninvasive modality for infant neuroimaging studies \cite{wang2019benchmark,chau2025scoping}.

\noindent 
Despite progress, robust hippocampal segmentation in infant MRI remains challenging. Classical atlas-based approaches often fail to generalize across heterogeneous populations and acquisition protocols \cite{schoemaker2016hippocampus,lidauer2022subcortical,sadil2024comparing}. Learning-based methods, including fully convolutional networks and U-Net variants \cite{wang2022id,zeng2020hippocampus,zhu2019dilated,li2023volumetric}, have improved performance but typically require large annotated datasets and extensive end-to-end training. These requirements are difficult to satisfy in infant neuroimaging, where expert annotations are scarce and expensive.

\noindent Foundation vision models pretrained on large-scale 2D natural images offer strong transferable representations. However, their direct application to volumetric medical data is non-trivial due to the inherent 3D structure of anatomy and the high memory cost of volumetric processing. Existing adaptation strategies often rely on fine-tuning or inserting trainable modules into the backbone, which increases complexity and reduces parameter efficiency. A key open question is how to leverage frozen 2D foundation encoders for 3D medical segmentation while preserving memory efficiency and generalization in low-data regimes.

\noindent In this work, we repurpose a frozen DINOv3~\cite{simeoni2025dinov3} Vision Transformer for volumetric hippocampus segmentation in infant MRI via a structured window-based disassembly--reassembly framework. Our approach converts slice-wise 2D features into volumetric representations with minimal trainable components and enables memory-aware training through sub-volume processing. 

\subsection*{Contributions}
Our contributions are threefold:
\begin{enumerate}
    \item We introduce a parameter-efficient framework that adapts a frozen 2D ViT to 3D medical segmentation by training only a lightweight dense prediction head.
    \item We propose a flexible sub-volume disassembly--reassembly strategy that enables linear memory scaling through independent fixed-size 3D windows.
    \item We demonstrate effective low-shot volumetric hippocampus segmentation in infant MRI, highlighting the potential of foundation models for data-scarce neuroimaging settings.
\end{enumerate}

\section{Related Work}

\textbf{Supervised Infant Hippocampus Segmentation.}
Early automated approaches relied on atlas-based pipelines constructed from adult brain MRI \cite{schoemaker2016hippocampus}. When applied to infant cohorts, these methods often exhibited systematic bias due to anatomical mismatch and protocol variability \cite{lidauer2022subcortical,sadil2024comparing}. Learning-based methods, particularly U-Net variants and fully convolutional architectures \cite{zhu2019dilated,zeng2020hippocampus,wang2022id,li2023volumetric}, have significantly improved segmentation accuracy. However, their performance typically depends on large-scale voxel-level annotations and extensive end-to-end optimization, which remain impractical in infant neuroimaging where expert labels are scarce and expensive to obtain.

\noindent \textbf{Foundation Models for Medical Segmentation.}
Large-scale vision foundation models pretrained on natural images have recently demonstrated strong transferability. Models such as SAM \cite{kirillov2023segment} provide powerful generic segmentation priors but are inherently designed for 2D imagery and prompt-based interaction. Subsequent efforts, including MedSAM \cite{MedSAM}, adapt these models to medical imaging through domain-specific fine-tuning. Nevertheless, direct deployment to volumetric neuroimaging remains challenging due to memory constraints, domain shift, and the structured 3D nature of anatomy.

\noindent \textbf{Parameter-Efficient Adaptation.}
To bridge the 2D--3D gap, recent works introduce parameter-efficient tuning mechanisms. Adapter-based approaches insert lightweight trainable modules into frozen backbones to support volumetric tasks \cite{gong20243dsam}, while methods such as SAMed employ LoRA-style updates for medical segmentation \cite{zhang2023customized}. Although computationally attractive, these strategies still modify internal backbone representations and require task-specific tuning. This can limit robustness in low-data regimes and reduces the plug-and-play flexibility of foundation encoders.

\noindent \textbf{Our Position.}
In contrast, our approach preserves a fully frozen 2D foundation encoder and instead performs structured volumetric reconstruction through window-based disassembly--reassembly with a lightweight decoder. This design targets memory efficiency, minimal parameter updates, and improved suitability for data-scarce infant MRI segmentation.
\section{Method}

\begin{figure}[ht]
    \centering
    \includegraphics[width=\textwidth]{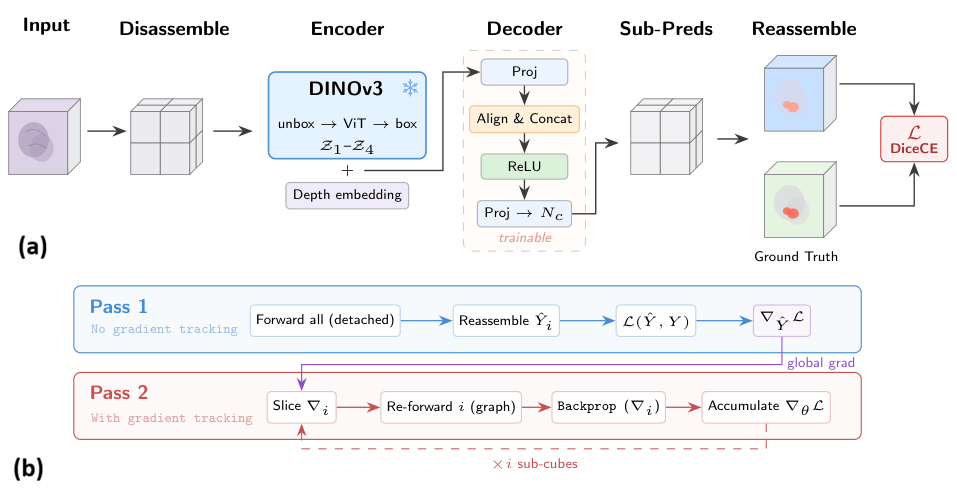}
    \caption{(a) A 3D MRI volume is decomposed into sub-cubes, encoded using a frozen 2D DINOv3 ViT with depth embedding, decoded via a lightweight head, and reassembled for global supervision. (b) Two-pass training: detached global loss computation followed by gradient-correct sub-cube backpropagation.}
    \label{fig:Figure1a-b-combined}
\end{figure}

Figure~\ref{fig:Figure1a-b-combined} illustrates the proposed framework for volumetric hippocampus segmentation in infant MRI. Our design adapts a pretrained 2D vision foundation model (VFM) to 3D medical data while keeping the backbone frozen. The framework consists of three components: (i) a slice-wise 3D-adapted DINOv3 encoder, (ii) a lightweight volumetric decoder, and (iii) a memory-aware sub-volume training strategy with two-pass gradient propagation.

\noindent \textbf{3D-Adapted Encoder Backbone.} We employ a pretrained DINOv3 ViT as a frozen feature extractor and adapt it to volumetric inputs through an \emph{unboxing--boxing} procedure. Given an input volume $\mathcal{X}\in\mathbb{R}^{1\times C\times D\times H\times W}$, we first unbox the volume into $D$ axial slices. Each slice is resized to the native DINOv3 resolution and processed independently by the frozen encoder, with no cross-slice interaction.

To capture hierarchical semantics, we extract intermediate token features from four transformer layers $\mathcal{L}=\{\ell_1,\ell_2,\ell_3,\ell_4\}$. These slice-wise tokens are then reassembled (boxing) into volumetric feature maps. For each level $\ell_k$, tokens from all slices are stacked and reshaped into
\[
\mathcal{F}_k \in \mathbb{R}^{1\times d_{\text{emb}}\times D\times (H/p)\times (W/p)}.
\]
To restore volumetric awareness, we add a learnable depth embedding that is interpolated when the input depth differs from the design depth. The encoder outputs the multi-scale volumetric features $\{\mathcal{F}_k\}_{k=1}^4$ for decoding.

\noindent \textbf{Lightweight Volumetric Decoder.} We design a parameter-efficient decoder inspired by DPT~\cite{ranftl2021vision} but simplified for volumetric efficiency. Each feature volume $\mathcal{F}_k$ is first projected using a $1\times1\times1$ convolution to reduce channel dimensionality. The projected features are refined using parallel $3\times3\times3$ convolutions to unify channel widths.

\noindent For multi-scale fusion, the shallowest feature defines the target spatial resolution via learned 3D transposed convolution. Deeper features are upsampled to this resolution and concatenated along the channel dimension. The fused representation is processed by two consecutive $3\times3\times3$ convolution blocks with instance normalization and ReLU activation to model local volumetric context. A final $1\times1\times1$ convolution produces the voxel-wise logits $\hat{\mathcal{Y}}$.

\noindent \textbf{Sub-volume Training Strategy.} Direct training on full 3D volumes is memory intensive. We therefore partition each volume into an even number of non-overlapping sub-cubes processed independently. To preserve global supervision while maintaining low memory usage, we introduce a two-pass gradient strategy.

In the first pass, all sub-cubes are forwarded without gradient tracking. Their predictions are detached and reassembled into the full-volume prediction $\hat{\mathcal{Y}}_{\text{full}}$, from which the global loss $\mathcal{L}(\hat{\mathcal{Y}}_{\text{full}}, \mathcal{Y}_{GT})$ is computed. Backpropagation at this stage yields the upstream gradient $\nabla_{\hat{\mathcal{Y}}_{\text{full}}}\mathcal{L}$ with respect to the full prediction.

In the second pass, each sub-cube is forwarded again with gradients enabled. For sub-cube $i$, we extract the corresponding gradient slice $\nabla_{\hat{\mathcal{Y}}_i}\mathcal{L}$ from the global gradient tensor and backpropagate it through the network. Gradients are accumulated across sub-cubes, followed by a single optimizer step. This procedure preserves exact global supervision while keeping the memory footprint bounded by the sub-cube size.

\noindent \textbf{Loss Function.} We optimize a composite Dice–Cross-Entropy loss to address severe class imbalance in hippocampus segmentation. The Dice term promotes region overlap, while the Cross-Entropy term provides stable voxel-wise gradients. The final objective is

\begin{equation}
\mathcal{L}_{\mathrm{DiceCE}}
= \lambda_{\mathrm{D}}\mathcal{L}_{\mathrm{Dice}}
+ \lambda_{\mathrm{CE}}\mathcal{L}_{\mathrm{CE}},
\end{equation}

where $\lambda_{\mathrm{D}}$ and $\lambda_{\mathrm{CE}}$ balance overlap accuracy and probabilistic supervision.

\section{Experiments and Results}

\subsection{Experiments}

\textbf{Dataset.}
We evaluate our method on the publicly available ALBERT Newborn Brain MRI dataset \cite{gousias2012magnetic}, which contains 20 infant subjects (15 preterm and 5 term) with expert annotations of subcortical structures. Each subject includes both T1- and T2-weighted scans; following prior work, we use the T2-weighted volumes for hippocampal segmentation.

\noindent \textbf{Preprocessing.}
Experiments are implemented in PyTorch with MONAI using binarized NIfTI volumes. All scans undergo a standardized pipeline: (1) reorientation to RAS, (2) resampling to a unified voxel spacing to reduce anisotropy, and (3) intensity standardization via percentile-based clipping followed by min–max scaling to $[0,1]$ and global z-score normalization. We extract a foreground-aware center crop of size $128\times128\times128$. During training only, we apply lightweight spatial augmentation including random axis flips and random $90^\circ$ rotations.

\noindent \textbf{Implementation Details.}
The encoder is a frozen DINOv3 ViT-Base backbone, while the decoder and depth embeddings are trainable, resulting in 21.3M learnable parameters. Models are trained for 100 epochs with batch size 1 using AdamW (initial learning rate $1\times10^{-4}$, weight decay $1\times10^{-4}$). We employ a linear warmup for the first 5 epochs followed by cosine annealing. Early stopping with patience 20 is applied based on validation Dice. Training uses the DiceCE loss with equal weighting, sigmoid activation, and background exclusion. Automatic mixed precision with gradient clipping (max\_norm=1.0) is used for stability and memory efficiency. We follow 5-fold cross-validation and report results from the first fold (80/20 split). All experiments are conducted on a single NVIDIA A40 (48\,GB) GPU.

\noindent \textbf{Evaluation Metrics.}
Performance is measured using Dice Similarity Coefficient (DSC), Intersection over Union (IoU), and volumetric error (\%), computed on the validation split.
\subsection{Results}

\noindent \textbf{Quantitative Evaluation.}
Table~\ref{tab:experiments_comparison} reports the performance on the ALBERT validation split. Training with the full $128^3$ volume (single sub-cube) substantially outperforms the decomposed setting. Specifically, the full-volume model achieves a DSC of 0.6514 and IoU of 0.4851, compared to 0.3518 DSC and 0.2148 IoU when the volume is partitioned into eight $64^3$ sub-cubes. A similar trend is observed for volumetric error, which is reduced from 30.42\% to 14.61\%.

\noindent This gap highlights the importance of preserving global spatial continuity for small-structure segmentation. While the proposed two-pass strategy enables memory-efficient training with sub-cubes, aggressive spatial fragmentation degrades performance, likely due to loss of long-range anatomical context and boundary consistency. Notably, the strong performance of the single-cube setting validates the effectiveness of the frozen DINOv3 features when sufficient volumetric context is retained.

\begin{table}[ht]
\centering
\caption{Comparison on the ALBERT dataset. The multi-cube setting partitions the volume into eight $64^3$ sub-cubes, while the single-cube setting processes the full $128^3$ crop. Results are reported for fold~1.}
\label{tab:experiments_comparison}

\setlength{\tabcolsep}{5pt}

\begin{tabular}{|l|c|c|c|}
\hline
\multirow{2}{*}{Experiment} & \multicolumn{3}{c|}{ALBERT (n=20)} \\
\cline{2-4}
 & DSC & IoU & Vol. Error (\%) \\
\hline
8 sub-cubes ($64^3$) & 0.3518 & 0.2148 & 30.42 \\
1 sub-cube ($128^3$) & \textbf{0.6514} & \textbf{0.4851} & \textbf{14.61} \\
\hline
\end{tabular}
\end{table}

\begin{figure}[ht]
    \centering
    \includegraphics[width=0.9\textwidth, height=0.25\textheight, keepaspectratio]{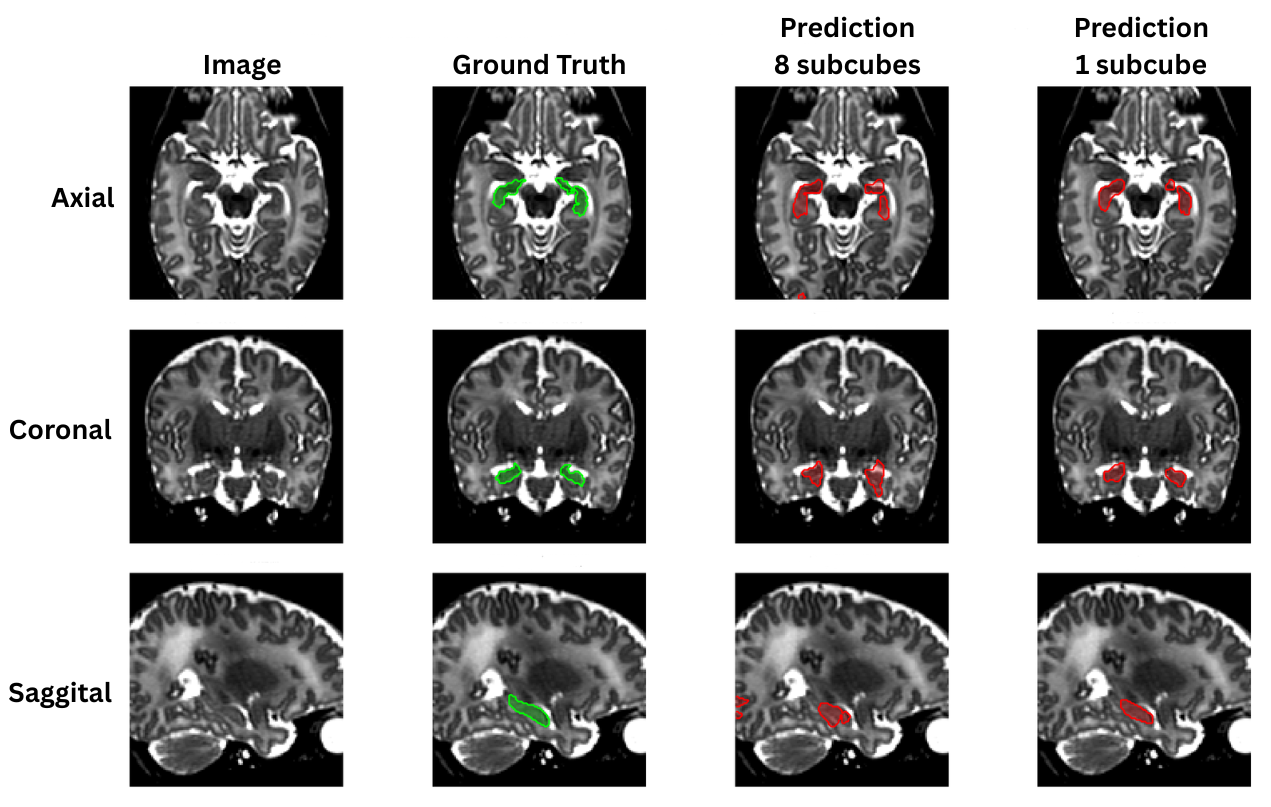}
    \caption{Qualitative comparison of hippocampus segmentation. The single-cube model shows improved anatomical coherence and boundary accuracy compared to the multi-cube setting.}
    \label{fig:Figure2-Eval Figures}
\end{figure}

\noindent \textbf{Qualitative Analysis.}
Figure~\ref{fig:Figure2-Eval Figures} provides representative validation visualizations across the three anatomical planes. The single-cube model produces more spatially coherent and anatomically faithful hippocampus boundaries, whereas the multi-cube setting exhibits fragmented predictions and boundary discontinuities. These artifacts are consistent with the quantitative degradation and suggest that excessive sub-volume partitioning weakens the model’s ability to capture global shape priors.

\noindent Overall, the results indicate that our framework benefits strongly from larger volumetric context, while the proposed sub-volume mechanism remains useful as a memory-control knob when full-volume training is infeasible.

\noindent \textbf{Ablation study.} We validated the contribution of two architectural components of our method: transformer layers and the learnable depth embedding. The experiments were conducted using the best-performing single-cube ($128^3$) configuration, with results for fold 1 reported in Table~\ref{tab:ablation_try}.

\noindent The results demonstrate that multi-scale feature fusion is critical for accurate segmentation. Simplifying to a single-scale head which only processes the deepest feature map from the ViT encoder, led to a decline in performance, with the DSC falling by about 45\% from the baseline. This highlights that fusing features from multiple transformer layers is essential for the decoding to reconstruct the fine-grained details and complex boundaries of the infant hippocampus.

\noindent In contrast, removing the depth embedding did not degrade performance; instead yielded a marginal improvement. It is hypothesized that its removal simplified the model slightly, in the full-volume context, possibly reducing minor overfitting.

\begin{table}[ht]
\centering
\caption{Ablation on the ALBERT dataset. These experiments are performed using the single-cube setting where a single full $128^3$ crop is processed. Results are reported for fold~1 only.}
\label{tab:ablation_try}

\setlength{\tabcolsep}{5pt}

\begin{tabular}{|l|c|c|c|}
\hline
\multirow{2}{*}{Experiment} & \multicolumn{3}{c|}{ALBERT (n=20)} \\
\cline{2-4}
 & DSC & IoU & Vol. Error (\%) \\
\hline
Baseline & 0.6514 & 0.4851 & 14.61 \\
No depth embedding & \textbf{0.6528} & \textbf{0.4872} & \textbf{11.71} \\
Single-scale decoding & 0.3585 & 0.2198 & 31.75 \\
\hline
\end{tabular}
\end{table}


Our results highlight a few insights about adapting a frozen 2D ViT for volumetric infant brain segmentation, for the hippocampus. The first being the importance of global spatial context. The drastic performance gap between the one single-3D window and eight-3D windows configuration implies that the decoding relies heavily on the full volumetric context of the hippocampus, rather than its sequential partitionings. This finding is particularly relevant for infant MRI, where the hippocampus occupies a small fraction already and its boundaries are inherently ambiguous due to low tissue contrast and myelination.

\noindent The ablation experiments highlight that multi-scale decoding is critical as removing it causes a substantial drop. This confirms that combining shallow edge and texture cues with deeper semantic features is critical towards accurate infant hippocampus segmentation. In contrast, depth embedding had a minimal impact, likely owing to 3D convolutions already capturing context in the single-subcube setting.

\noindent A broader contribution of our work is that frozen foundation models’ backbones, pretrained on 2D natural images, can serve as effective feature extractors for 3D medical imaging even without any encoder fine-tuning. Our method yielding a segmentation performance with only a lightweight trainable decoder on a dataset of 20 cases is encouraging for many low-resource neuroimaging scenarios where annotated data is scarce and limited.

\section{Conclusion}

We presented a parameter-efficient framework for volumetric hippocampus segmentation in infant MRI by repurposing a frozen 2D DINOv3 Vision Transformer. Our approach bridges the 2D-3D gap through slice-wise encoding, depth-aware volumetric reconstruction, and a lightweight 3D decoder, while maintaining minimal trainable components. To address memory constraints inherent in volumetric learning, we introduced a sub-volume disassembly--reassembly strategy with two-pass gradient propagation that preserves global supervision under bounded memory.

\noindent Experiments on the ALBERT dataset demonstrate that retaining full volumetric context is critical for accurate small-structure segmentation. While the proposed sub-volume mechanism enables scalable training, aggressive spatial partitioning degrades anatomical fidelity, highlighting the importance of global continuity in infant hippocampus delineation. Overall, the results validate the effectiveness of frozen foundation features for data-scarce neuroimaging when paired with appropriate volumetric adaptation.

\noindent Future work will explore context-aware sub-volume fusion, cross-dataset generalization, and extension to multi-structure infant brain segmentation, further advancing parameter-efficient use of vision foundation models in 3D medical imaging.

%
%
%
\bibliographystyle{splncs04}
\bibliography{arxiv}
%




\end{document}